\documentclass[aps,prd,twocolumn,superscriptaddress,10pt,floatfix,nofootinbib,showpacs,preprintnumbers,amsmath,amssymb]{revtex4-2}
\usepackage[utf8]{inputenc}
\usepackage{amsfonts,bm,graphicx,booktabs,array,xcolor}
\usepackage[colorlinks=true,allcolors=blue]{hyperref}

\newcommand{\dd}{\mathrm d}
\newcommand{\Sch}{\mathrm{Sch}}
\newcommand{\B}{\mathrm B}
\newcommand{\rin}{r_{\rm in}}
\newcommand{\rmbo}{r_{\rm MBO}}
\newcommand{\bc}{b_{\rm c}}
\newcommand{\chitr}{\chi_{\rm tr}}
\providecommand{\doi}[1]{\href{https://doi.org/#1}{doi:\nolinkurl{#1}}}

\begin{document}
\title{Can absorption enhance differences between black-hole images?}
\author{Tong Li}
\affiliation{Department of Physics, School of Science, Jiangsu University of Science and Technology, Zhenjiang 212003, People's Republic of China}
\author{Jian-Yun Wan}
\affiliation{Department of Physics, School of Science, Jiangsu University of Science and Technology, Zhenjiang 212003, People's Republic of China}
\author{Zi-Liang Wang}
\email{ziliang.wang@just.edu.cn}
\affiliation{Department of Physics, School of Science, Jiangsu University of Science and Technology, Zhenjiang 212003, People's Republic of China}
\affiliation{Theoretical Physics Research Center, School of Science, Jiangsu University of Science and Technology, Zhenjiang 212003, People's Republic of China}
\date[]{}
\begin{abstract}
We investigate whether absorption can enhance morphological differences between images of different spacetimes, particularly when their critical impact parameters are similar. Using a phenomenological thick-accretion-flow model with an effective absorption coefficient $\chi_0$,  we compare Schwarzschild black-hole images with those of Bardeen, Hayward, and Ay\'on-Beato--Garc\'ia black holes. In the face-on Schwarzschild--Bardeen comparison with $g/M=0.5$, the dominant brightness shifts from a narrow lensed feature to a broader outer feature at $\chi_0M\simeq0.585$ and $0.717$, respectively. Near the first transition, the integrated difference between independently normalized images increases by $23.5\%$ on a common physical grid and by $46.6\%$ after rescaling by the respective critical impact parameters. Enhancement occurs across all three regular-black-hole families and over a broad range of the sampled flow parameters and viewing inclinations, although its magnitude and occurrence depend on the prescribed flow. Asynchronous brightness-peak transitions alone are insufficient to guarantee enhancement.  These results suggest that the opacity dependence of image morphology may provide information complementary to the critical-curve scale, potentially helping distinguish spacetimes with similar characteristic shadow sizes. Assessing this possibility observationally requires physically consistent, frequency-dependent radiative transfer.
\end{abstract}
\maketitle

\section{Introduction}\label{sec:introduction}

A black-hole image carries information about both the underlying spacetime geometry and the surrounding emitting matter.  The images of M87* and Sagittarius A* obtained
by the Event Horizon Telescope (EHT) have made this connection accessible to
observation~\cite{EHT2019I,EHT2022I}. In particular, the 2017 observations of  M87* at 1.3 mm (230 GHz) revealed a bright ring with an angular diameter of
$42\pm3\,\mu\mathrm{as}$ surrounding a central brightness depression.
At this wavelength, the plasma is sufficiently transparent for strong
gravitational lensing of emission near the black hole to be
visible~\cite{EHT2019I}.

The measured ring diameter is consistent with the characteristic angular
scale of a black-hole shadow~\cite{EHT2019I}. For a Schwarzschild black hole, the shadow boundary is determined by the unstable photon sphere at $r_{\rm ph}=3M$, corresponding to a critical impact parameter $\bc=3\sqrt{3}M$ for a distant observer~\cite{PerlickTsupko2022}. The EHT results have stimulated
extensive studies of black-hole shadows~\cite{Gralla2019,Psaltis2020,Vagnozzi2023,Wang:2023rjl,Vincent2022} and tests of spacetime geometry~\cite{Zare:2024dtf,Wang:2025fmz,Heidari:2024bbd,Errehymy:2026vhl,Yang:2025whw,Li:2026vpa,Wang:2025vsx,Xiong:2025wgs,Lim:2025cne,Zhang:2026ttf,Ahmed:2025sav},
with particular attention to the shadow size and its relation to the critical
impact parameter.

A larger ring was subsequently resolved at a longer wavelength.
Using 2018 observations at 3.5 mm (86 GHz) with the Global Millimetre VLBI
Array (GMVA), supplemented by the Atacama Large Millimetre/submillimetre
Array (ALMA) and the Greenland Telescope (GLT), Lu et al.~\cite{Lu2023}
reported a ring diameter of $64^{+4}_{-8}\,\mu\mathrm{as}$, approximately
$50\%$ larger than the 1.3 mm ring. For the mass and distance adopted
in that study, this corresponds to an apparent ring radius of
$8.4^{+0.5}_{-1.1}M$ in geometric units. The ring was also broader. In the general relativistic magnetohydrodynamic (GRMHD)-based radiative-transfer models of Lu et al.~\cite{Lu2023}, the plasma around the black hole is optically thin at 1.3 mm but becomes optically thick at 3.5 mm owing to synchrotron self-absorption. Their thermal synchrotron model reproduces the larger and broader ring observed at the longer wavelength, highlighting the role of absorption in shaping the observed emission structure. These findings provide a concrete motivation for investigating how absorption can modify the apparent ring size and brightness distribution without changing the underlying spacetime geometry.

The interpretation of black-hole images is complicated by the dependence of their brightness distributions on the accretion flow. In simple optically thin models, the shadow boundary remains closely tied to the underlying photon sphere~\cite{Narayan2019}, whereas the location, width, and relative brightness of emission features depend on the distribution of emitting matter~\cite{Gralla2019}. For regular black holes, different radial emissivity prescriptions can alter the relative contributions of direct and lensed emission~\cite{DeMartino2023}, while uncertainties in the accretion flow can be degenerate with changes in the spacetime geometry~\cite{Lara2021}. 

Frequency-dependent black-hole imaging and the transition between optically thin and optically thick emission have been investigated using numerical and analytical accretion-flow models~\cite{Chael2023,Desire2025}. In particular, an effective-absorption model for a geometrically thick flow in Schwarzschild spacetime has demonstrated substantial changes in image morphology between weakly and strongly attenuated regimes~\cite{Wang2025}. However, a change in the appearance of an individual black-hole image does not necessarily imply an increased difference between images of distinct spacetimes. This motivates the central question of the present work: Can absorption enhance morphological differences between black-hole images, even when their critical impact parameters are similar? If the dominant brightness features switch at different opacities in two spacetimes, does this necessarily amplify the overall difference between their images?

A self-consistent comparison across observing bands would require modeling the
plasma dynamics, thermodynamics, and magnetic structure, together with the
frequency dependence of emission and absorption. A standard approach combines
GRMHD simulations with relativistic
radiative transfer~\cite{EHT2019V,Mizuno2018,Pelle2022}. Building such a model library
across several metrics and flow parameters requires substantial computational
resources. We instead use a simplified effective model, varying absorption
while keeping the emission prescription fixed. This allows us to examine
whether absorption can enhance differences between spacetime images before
attempting a self-consistent multifrequency calculation. 

We consider static, spherically symmetric black-hole geometries, comparing Schwarzschild with Bardeen, Hayward, and Ay\'on--Beato--Garc\'ia (ABG) spacetimes~\cite{AyonBeato2000,Hayward2006,AyonBeato1998,WangBattista2026}. The accretion flow is described by prescribed geometry, velocity, emissivity, and effective absorption, with the same modeling prescription applied to each spacetime. The principal example is a face-on Bardeen model with $g/M=0.5$.  We quantify the differences using two integrated measures: one compares images on a common physical impact-parameter grid, while the other rescales each image by its own critical impact parameter to remove the overall critical-curve scale difference.

We find that absorption can enhance the differences between black-hole images over an intermediate opacity range. By tracking the narrow and broad brightness features, we examine how their relative evolution contributes to the integrated image differences. We then investigate the dependence of this enhancement on the spacetime geometry, flow velocity, inner boundary, thickness, and observer inclination. Importantly, asynchronous changes in the dominant brightness features do not necessarily guarantee an enhancement of the integrated image differences. Our results identify a conditional absorption-induced morphological effect and provide a starting point for future comparisons based on more realistic radiative-transfer models.

The paper is organized as follows. Section~\ref{sec:model} introduces the spacetime geometries, accretion-flow model, and radiative-transfer prescription. The baseline image evolution and absorption-induced image differences are presented in Sec.~\ref{sec:phenomenon}, followed by an examination of their dependence on spacetime geometry and accretion-flow parameters in Sec.~\ref{sec:conditions}. The main findings and limitations are discussed in Sec.~\ref{sec:discussion}. Finally, Appendix~\ref{app:optical-depth} examines the role of optical depth in brightness-peak exchanges, while Appendix~\ref{app:inner-radius} explores the dependence of the results on the flow's inner boundary.

\section{Geometry and accretion-flow model}\label{sec:model}
\subsection{Background metrics and characteristic radii}

We work in geometric units $G=c=1$ and consider black-hole spacetimes with the same ADM mass $M$. Their line elements take the form
\begin{equation}
 \dd s^2=-f(r)\dd t^2+f(r)^{-1}\dd r^2+
 r^2(\dd\theta^2+\sin^2\theta\,\dd\varphi^2),
 \label{eq:metric}
\end{equation}
where the metric functions for the Schwarzschild and Bardeen spacetimes are given by
\begin{equation}
 f_{\Sch}(r)=1-\frac{2M}{r},\qquad
 f_{\B}(r)=1-\frac{2Mr^2}{(r^2+g^2)^{3/2}}.
 \label{eq:metric-functions}
\end{equation}
For the Hayward and Ayón--Beato--García (ABG) spacetimes, the corresponding metric functions read
\begin{equation}
 \begin{split}
 f_{\rm H}(r)&=1-\frac{2Mr^2}{r^3+2M\ell^2},\\
 f_{\rm ABG}(r)&=1-\frac{2Mr^2}{(r^2+q^2)^{3/2}}
 +\frac{q^2r^2}{(r^2+q^2)^2}.
 \end{split}\label{eq:additional-metrics}
\end{equation}
Here $g$ is the Bardeen regularization parameter, which can be interpreted as a magnetic charge in a nonlinear-electrodynamics realization~\cite{AyonBeato2000}, $\ell$ is the characteristic length scale governing the regular core of the Hayward spacetime~\cite{Hayward2006}, and $q$ is the electric charge parameter of the ABG solution~\cite{AyonBeato1998}. Throughout this work, we use the dimensionless parameters $g/M$, $\ell/M$, and $q/M$ to characterize the respective black-hole families.

The outer event horizon is located at $r_h$, defined as the largest positive root of $f(r_h)=0$. The photon-sphere radius $r_{\rm ph}$ corresponds to the outermost unstable circular null orbit, where the effective potential for null geodesics has a local maximum. The photon-sphere radius and the associated critical impact parameter $b_c$ are determined by
\begin{equation}
 r_{\rm ph}f'(r_{\rm ph})-2f(r_{\rm ph})=0,\qquad
 \bc=\frac{r_{\rm ph}}{\sqrt{f(r_{\rm ph})}}.
 \label{eq:photon-sphere}
\end{equation}
Here and throughout, a prime denotes differentiation with respect to $r$.
For a test particle moving along an equatorial timelike geodesic, the conserved specific energy and angular momentum are defined by \(E=-u_t\) and \(L=u_\phi\), respectively. For circular orbits, they are given by
\begin{equation}
 E^2=\frac{2f^2}{2f-rf'},\qquad
 L^2=\frac{r^3f'}{2f-rf'}.
 \label{eq:timelike}
\end{equation}
The marginally bound circular orbit (MBO) is determined by $E(r_{\rm MBO})=1$. The innermost stable circular orbit (ISCO) is identified by the marginal stability condition $V_{\rm eff}''(r_{\rm ISCO})=0$, which yields
\begin{equation}
rff''+3ff'-2r(f')^2=0,
\label{eq:isco}
\end{equation}
where the physically relevant root is selected on the admissible circular-orbit branch.

The relevant radii for Schwarzschild geometry are $(r_h,r_{\rm ph},\rmbo,r_{\rm ISCO})=
(2,3,4,6)M$, with $\bc=3\sqrt{3}M$. The baseline Bardeen model has
$g/M=0.5$ and
\begin{equation}
 \begin{gathered}
 (r_h,r_{\rm ph},\rmbo,r_{\rm ISCO},\bc)/M\\
 =(1.78597,2.76871,3.69306,5.57222,4.96004).
 \end{gathered}
\end{equation}
The critical impact parameter of the baseline Bardeen spacetime is approximately $4.5\%$ smaller than its Schwarzschild counterpart. This deviation lies within the $68\%$ allowed range inferred from the EHT observations of Sgr A* and thus cannot be ruled out by the shadow-size constraint alone~\cite{EventHorizonTelescope:2022xqj}.



All photons in this work follow null geodesics of the background metric in
Eq.~\eqref{eq:metric}. This is a test-photon prescription. If a regular
metric is instead interpreted as a particular nonlinear-electrodynamics
solution, electromagnetic perturbations may propagate on an effective
optical metric~\cite{Walia2024}. That distinct problem is not addressed here.

\subsection{Flow geometry, velocity, and emissivity}

\begin{figure*}[t!]
 \centering
 \includegraphics[
   width=0.98\textwidth
 ]{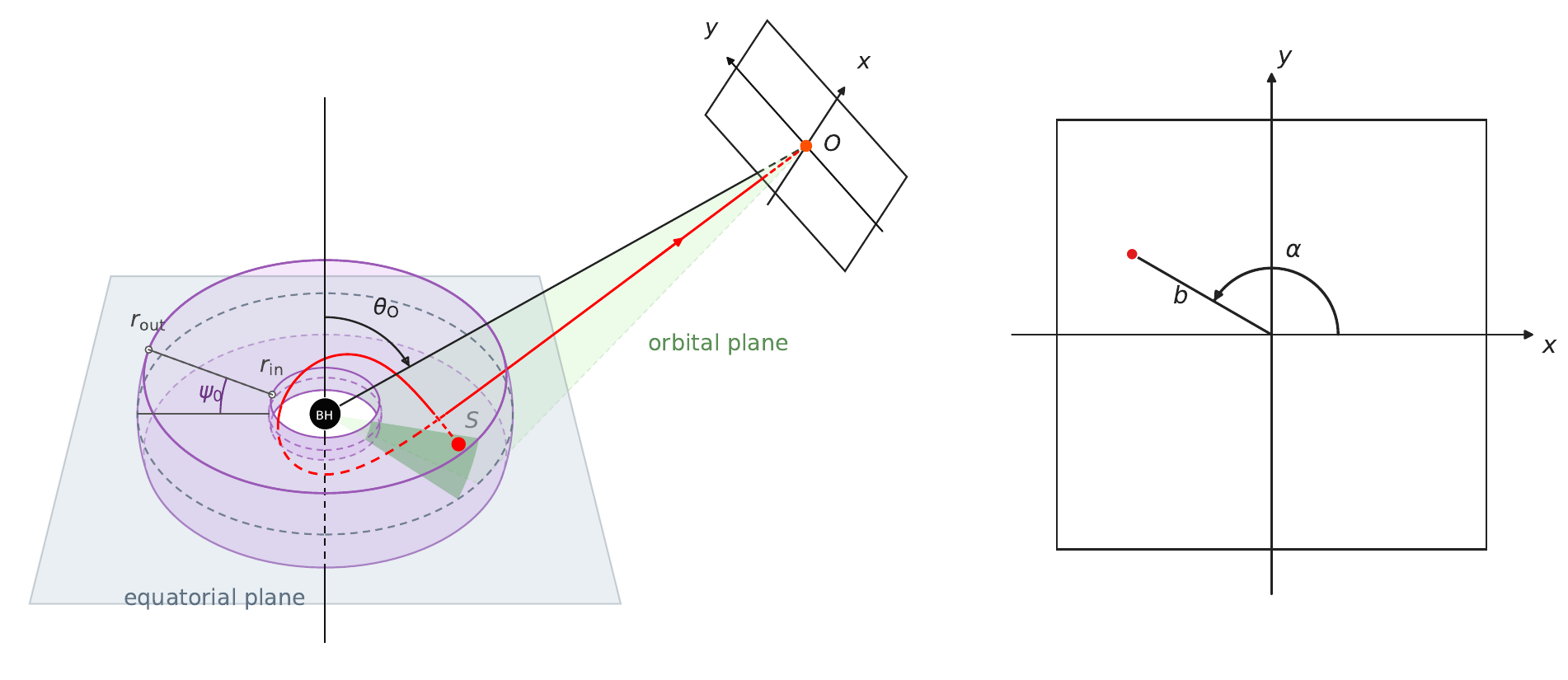}
\caption{Schematic illustration of the geometrically thick accretion-flow model and the observer's image plane. Left panel: The emitting and absorbing region extends from $r_{\rm in}$ to $r_{\rm out}$, with a half-opening angle $\psi_0$ measured from the equatorial plane. The observer $O$ is located at an inclination angle $\theta_O$ relative to the symmetry axis. A representative photon trajectory (red) lies in the orbital plane (green). Right panel: The observer's image-plane coordinates, where each ray is characterized by its impact parameter $b$ and azimuthal angle $\alpha$. }
 \label{fig:disk-sketch}
\end{figure*}

The flow occupies an axisymmetric wedge about the equatorial
plane, defined by $|\psi|\leq\psi_0$, where
$\psi=\pi/2-\theta$ and $\psi_0$ is the half-opening angle.
The emitting and absorbing region extends radially from
$r_{\rm in}$ to $r_{\rm out}=50M$. The inner boundary $r_{\rm in}$ also enters
the radial emissivity prescribed below. Unless otherwise specified, we adopt $r_{\rm in}=r_{\rm MBO}$ for each background metric as a phenomenological prescription, motivated by the image morphology discussed below and further examined in Appendix~\ref{app:inner-radius}.
The observer inclination angle $\theta_O$ is measured
from the symmetry axis of the flow. The geometry of the adopted accretion flow is schematically illustrated in Fig.~\ref{fig:disk-sketch}.

The flow velocities are prescribed phenomenologically as
\begin{equation}
v^r\equiv\frac{\dd r}{\dd t}
=-\kappa_{\rm ff}f\sqrt{1-f},
\qquad
\Omega\equiv\frac{\dd\varphi}{\dd t}
=\kappa_K\sqrt{\frac{f'}{2r}}.
\label{eq:velocity}
\end{equation}
 The dimensionless parameters $\kappa_{\rm ff}$ and $\kappa_K$ control the radial infall and azimuthal rotation, respectively. The radial velocity is scaled relative to that of a particle freely falling from rest at infinity, while the angular velocity is scaled relative to the Keplerian angular velocity of equatorial circular geodesics. The corresponding four-velocity is $u^\mu=u^t(1,v^r,0,\Omega)$, where the normalization condition $u^\mu u_\mu=-1$ gives
\begin{equation}
(u^t)^{-2}
=f-\frac{(v^r)^2}{f}
-r^2\cos^2\psi\,\Omega^2.
\label{eq:ut}
\end{equation}
Only configurations satisfying the timelike condition $(u^t)^{-2}>0$ are considered.

We adopt an effective radial emissivity based on the viscous-dissipation prescription of the Shakura--Sunyaev thin-disk model~\cite{Frank2002}, phenomenologically extended to geometrically thick accretion flows in Ref.~\cite{Wang2025}:
\begin{equation}
j(r)=\frac{\Omega'(r)}{4\pi\psi_0}
\left[-\Omega(r)+\frac{\rin^2\Omega(\rin)}{r^2}\right],
\qquad \rin\leq r\leq50M.
\label{eq:emissivity}
\end{equation} Here $\psi_0$ is in radians and the wedge restriction is understood.

For the adopted accretion-flow configuration, the resulting images can exhibit two distinct brightness features: a narrow ring near the critical curve in the optically thin limit and a broader emission component at larger impact parameters that can become dominant as absorption increases.  This behavior is qualitatively motivated by the wavelength-dependent ring morphology observed in M87*. In particular, the ring observed at $3.5$ mm is larger and broader than that at $1.3$ mm, with a measured diameter of $8.4^{+0.5}_{-1.1}$ Schwarzschild radii~\cite{Lu2023}, corresponding to an apparent radius of approximately $7.3M$--$8.9M$.

As will be shown in Sec.~\ref{sec:phenomenon}, our numerical calculations for the Schwarzschild baseline exhibit a dominant outer brightness peak at $b\simeq 7$--$8M$ over a finite range of absorption coefficients, with its position falling within the observational reference range for part of this interval. Among the characteristic inner-radius prescriptions tested, $\rin=\rmbo$ permits both a narrow ring near $b_{\rm c}$ in the optically thin limit and a larger, broader ring at finite absorption, whereas the photon-sphere and ISCO choices do not produce a dominant outer brightness peak within the observational reference range when the other baseline parameters are held fixed.

We emphasize that this qualitative agreement serves only as a morphological motivation for the adopted phenomenological model, rather than a quantitative fit to the EHT observations or a self-consistent frequency-dependent emission model. The dependence of the results on the inner-radius prescription is further examined in Appendix~\ref{app:inner-radius}.

Unless otherwise specified, we adopt the following baseline parameters in our numerical calculations:
\begin{equation}
 \begin{gathered}
 \theta_O=0^\circ,\qquad \psi_0=10^\circ,\\
 (\kappa_{\rm ff},\kappa_K)=(0.1,0.9),\qquad
 \rin=\rmbo.
 \end{gathered}
 \label{eq:baseline-parameters}
\end{equation}
The inner radius $r_{\rm MBO}$ is evaluated separately for each spacetime.

\subsection{Ray tracing and effective absorption}\label{sec:ray-tracing}

In our numerical calculations, null rays are traced backward from a distant observer located at $r_O=10^5M$.\footnote{This approximation is appropriate for the asymptotically flat spacetimes considered here. In non-asymptotically flat spacetimes, however, the observed angular size of the black-hole shadow can depend sensitively on the observer's position and state of motion~\cite{Perlick:2018iye,Li:2026amn}.} Each ray is characterized by its impact parameter $b$ and azimuthal angle $\alpha$ on the observer's image plane, as illustrated in Fig.~\ref{fig:disk-sketch}. The corresponding Cartesian coordinates are defined by $x=b\cos\alpha$ and $y=b\sin\alpha$, with the origin at the projected black-hole center. Owing to spherical symmetry, each photon trajectory lies in a fixed orbital plane. We denote the polar angle in this plane, measured about the black-hole center, by $\phi$, to distinguish it from the spacetime azimuthal coordinate $\varphi$. With $\mu=1/r$, its orbit equation is
\begin{equation}
 \frac{\dd^2\mu}{\dd\phi^2}=-\mu f(1/\mu)+\frac12 f'(1/\mu).
 \label{eq:ray}
\end{equation}
Let $\lambda$ denote an affine parameter along the null geodesic. Setting the photon energy at infinity to unity gives 
\begin{equation}
 \dd\lambda=\frac{r^2}{b}\dd\phi.
 \label{eq:affine}
\end{equation}

For the distant static observer, the redshift is
\begin{equation}
 g_{\rm red} =\frac{1}{u^t[1+v^r(k_r/k_t)+\Omega(k_\varphi/k_t)]}.
 \label{eq:redshift}
\end{equation}

Following the phenomenological radiative-transfer prescription of Ref.~\cite{Wang2025}, we express the unattenuated contribution to the observed bolometric intensity as
\begin{equation}
\dd I_0=j(r)g_{\rm red}^4\,\dd\lambda,
\end{equation}
where $j(r)$ denotes the effective radial emissivity per unit coordinate volume introduced in Eq.~\eqref{eq:emissivity}.\footnote{The proper-volume correction $\sqrt{f(r)}$ cancels the corresponding factor in the local static-frame
path length, yielding $j_{\rm proper}\dd\ell_{\rm stat}
=j(r)\dd\lambda$ for $E=1$.} 
The effective absorption is prescribed independently through
\begin{equation}
\dd\tau=\chi_0\,\dd\ell_{\rm stat},
\qquad
\dd\ell_{\rm stat}=\frac{\dd\lambda}{\sqrt{f}},
\label{eq:optical-depth}
\end{equation}
where $\tau$ denotes the dimensionless optical depth, $\chi_0$ is a constant effective absorption coefficient, and $\dd\ell_{\rm stat}$ is the infinitesimal photon path length measured by a local static observer. Related constant-coefficient approximations have been employed in black-hole imaging and optical-depth calculations~\cite{Fukue2011,Hou2022KerrMelvin,Zheng:2024ftk,Desire2025}.

Assuming that the effective source contribution per unit optical depth is approximately constant within each integration cell, the radiative-transfer equation gives
\begin{equation}
 I_{\rm out}=I_{\rm in}e^{-\Delta\tau}
 +\Delta I_0\frac{1-e^{-\Delta\tau}}{\Delta\tau},
 \label{eq:cell}
\end{equation}
where $\Delta I_0$ is the unattenuated intensity contribution from the cell and $\Delta\tau$ is its optical depth. The factor $(1-e^{-\Delta\tau})/\Delta\tau$ approaches unity as $\Delta\tau\to0$. Although the null geodesics are traced backward from the observer, the radiative transfer is evaluated in the forward direction, from the far end of each ray toward the observer.


Let $I_a(b,\alpha;\chi_0)$ denote the observed bolometric intensity in spacetime $a$. To compare the relative brightness distributions, we normalize each image by its own maximum, a convention commonly adopted in theoretical black-hole imaging studies~\cite{Younsi2023,likhachev2022high,Vincent2022,Mizuno2018,Eichhorn2023,Ricarte:2014nca,Chen:2024ilc,Fauzi:2024nta,BoosHu2026}. Specifically, at each value of $\chi_0$, we define
\begin{equation}
 \widehat I_a(b,\alpha;\chi_0)=
 \frac{I_a(b,\alpha;\chi_0)}{\max_{b,\alpha}I_a(b,\alpha;\chi_0)}.
 \label{eq:peak-normalization}
\end{equation}
This normalization removes the overall intensity scale while preserving the relative brightness of different image features. For a face-on observer, axial symmetry implies $I_a(b,\alpha;\chi_0)=I_a(b;\chi_0)$, so the corresponding radial profiles are described by $\widehat I_a(b;\chi_0)$. Unless otherwise specified, all image panels, radial profiles, and image-difference measures presented below employ the normalization in Eq.~\eqref{eq:peak-normalization}.



\section{Absorption-driven changes in black-hole images}\label{sec:phenomenon}
\subsection{A face-on Schwarzschild--Bardeen example}

\begin{figure*}[t!]
 \centering
 \includegraphics[width=0.995\textwidth]{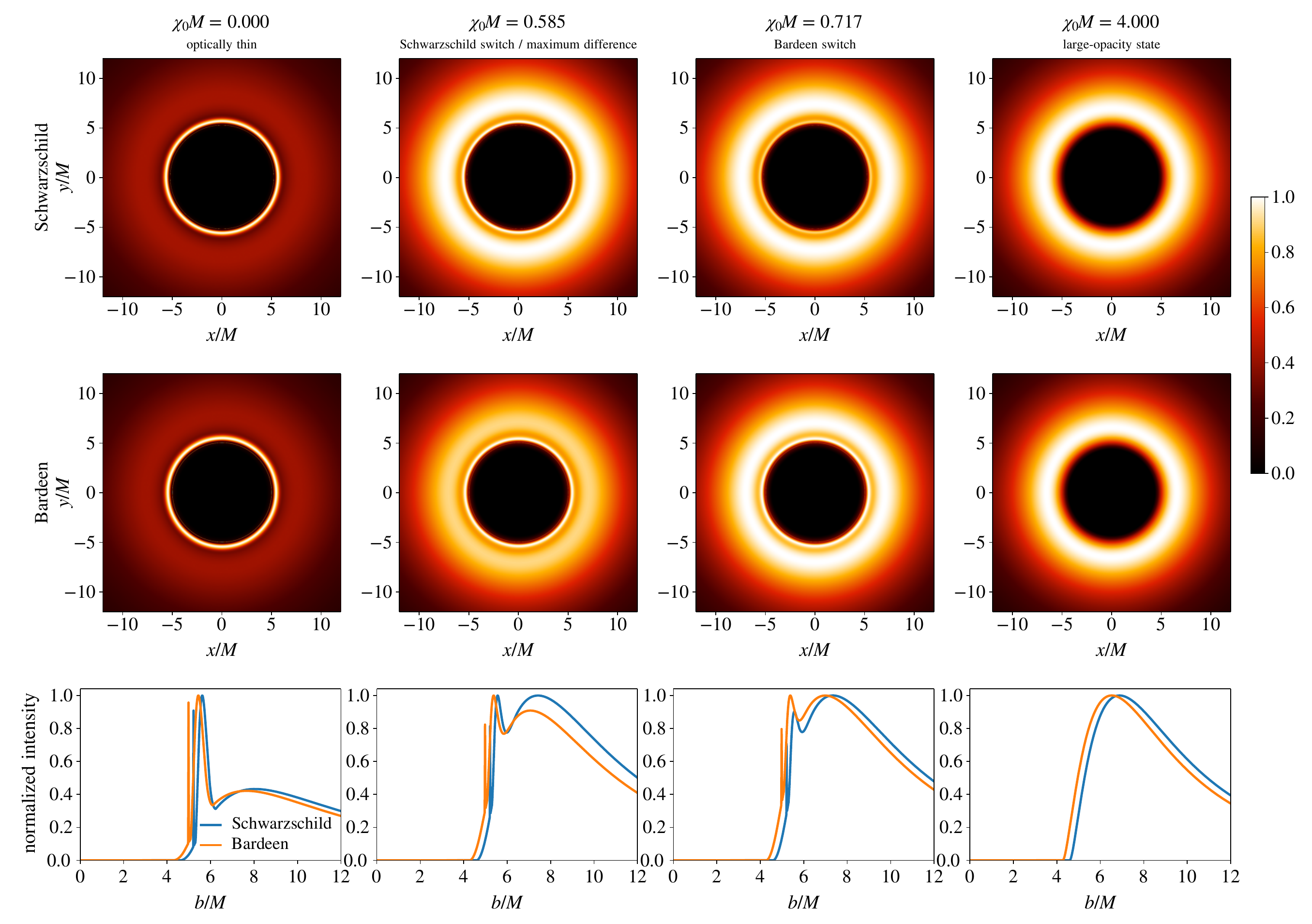}
 \caption{Evolution of black-hole images and radial intensity profiles with increasing absorption for the face-on Schwarzschild and Bardeen spacetimes. The model parameters are $g/M=0.5$, $\psi_0=10^\circ$, $\rin=\rmbo$, and $(\kappa_{\rm ff},\kappa_K)=(0.1,0.9)$. The first two rows show the Schwarzschild and Bardeen images, respectively, while the third row compares their radial intensity profiles. From left to right, the columns correspond to $\chi_0M=0$, $0.585$, $0.717$, and $4$, representing the optically thin limit, the Schwarzschild peak transition, the Bardeen peak transition, and a strongly attenuated state, respectively. Each image is independently normalized by its maximum intensity, and the same linear color scale is used throughout. The narrow lensed feature dominates in the optically thin limit, whereas the broad outer feature becomes dominant at larger absorption, with the transition occurring at different opacities in the two spacetimes.}
 \label{fig:baseline}
\end{figure*}

Figure~\ref{fig:baseline} illustrates how the Schwarzschild and Bardeen images evolve with increasing absorption. We adopt the face-on baseline parameters in Eq.~\eqref{eq:baseline-parameters}, with $g/M=0.5$ for Bardeen. The four columns correspond to the optically thin limit, the brightness-peak transitions in Schwarzschild and Bardeen, respectively, and a strongly attenuated state.

In the absence of absorption, both radial intensity profiles exhibit two local maxima: a narrow peak just outside the respective critical impact parameter $b_{\rm c}$ and a broader peak at larger impact parameters. Following the terminology of direct and lensed emission discussed in Ref.~\cite{Gralla2019}, we refer to these as the narrow lensed and broad direct-emission features, respectively.  Initially, the narrow feature dominates both images. As absorption increases, the broad feature becomes progressively brighter relative to the narrow one and eventually dominates.

This behavior can be qualitatively attributed to the
different absorbing paths sampled by the two features.
The narrow lensed feature is more strongly attenuated
because its emission typically encounters larger foreground
optical depths, whereas the broad direct-emission feature
receives a greater contribution from less attenuated
foreground radiation. This interpretation is supported by
the optical-depth analysis in
Appendix~\ref{app:optical-depth}.

The transition from narrow-peak to broad-peak dominance
occurs at different absorption coefficients in the two
spacetimes. We define the transition opacity $\chitr$
as the value at which the narrow and broad local brightness
maxima become equal. This criterion characterizes the
relative brightness of the two features rather than a
global transition from optically thin to optically thick
emission. For the baseline configuration, we obtain

\begin{equation}
 \chitr^{\Sch}M\simeq0.585,\qquad
 \chitr^{\B}M\simeq0.717.
 \label{eq:transition-values}
\end{equation}

At the first transition, the broad Schwarzschild peak
near $b\simeq7.4M$ becomes as bright as its narrow
counterpart, whereas the broad peak in Bardeen remains
fainter. This contrast is evident in the second column
of Fig.~\ref{fig:baseline}, as well as in the corresponding
radial intensity profiles. Within the interval
$0.585<\chi_0M<0.717$, the Schwarzschild image is
dominated by the broad peak, while the Bardeen image
remains narrow-peak dominated. At $\chi_0M\simeq0.717$,
the two peaks in Bardeen become equally bright.
In the strongly attenuated state at $\chi_0M=4$,
both images are dominated by their broad peaks,
although their locations and widths remain different.

The different transition opacities arise from the distinct
responses of the relative peak brightnesses to absorption
in the two spacetimes. Although the same effective
absorption coefficient is adopted, differences in photon
trajectories, gravitational redshift, and the metric-dependent
inner boundary and emissivity affect the relative attenuation
of the narrow and broad features. These effects cause
the narrow-to-broad peak intensity ratio to reach unity
earlier in Schwarzschild than in Bardeen.
The optical-depth analysis in
Appendix~\ref{app:optical-depth} provides further insight
into this behavior, showing that the peak exchange
depends on the distribution of foreground absorption
rather than on the total ray optical depth alone.

These results demonstrate that absorption can alter
the dominant brightness features of black-hole images. Such morphological
differences are not captured by the critical impact
parameter $b_{\rm c}$ alone. Whether they lead to an
enhancement of the overall image difference is examined
quantitatively in the following subsection.

\subsection{Quantifying image differences}
\label{sec:diagnostics}

The image sequence in Fig.~\ref{fig:baseline} raises two quantitative questions: can absorption enhance the differences between black-hole images relative to the optically thin limit, and does this enhancement persist after accounting for the different critical-curve scales? To address these questions, we introduce two integrated measures of the differences between normalized images. These measures provide a quantitative basis for comparing different absorption coefficients and model parameters, while the images, radial intensity profiles, and radial contributions help identify the features responsible for the observed changes.

The normalized $L_1$ difference\footnote{The $L_1$ norm measures the integrated absolute value of a function. Here, it quantifies the total brightness difference between two images.} on the common physical grid is
\begin{widetext}
\begin{equation}
D_{\rm phys}(\chi_0)=
\frac{\displaystyle\int_{\cal A}
|\widehat I_{\Sch}(b,\alpha;\chi_0)-\widehat I_{\rm R}(b,\alpha;\chi_0)|
\,\dd A}
{\displaystyle\frac12\int_{\cal A}
[\widehat I_{\Sch}(b,\alpha;\chi_0)+\widehat I_{\rm R}(b,\alpha;\chi_0)]
\,\dd A},
\qquad \dd A=b\,\dd b\,\dd\alpha.
\label{eq:dphys}
\end{equation}
\end{widetext}
Here the subscript R denotes the regular black-hole spacetime under comparison, and $\cal A$ is the common image-plane integration domain. The area element $\dd A=b\,\dd b\,\dd\alpha$ accounts for the increasing image area at larger impact parameters. The resulting dimensionless measure satisfies $0\leq D_{\rm phys}\leq2$, with $D_{\rm phys}=0$ for identical normalized images. For face-on observations, azimuthal symmetry reduces the expression to a one-dimensional radial integral with measure $b\,\dd b$.

To identify the radial regions responsible for the integrated image difference, we define the contribution per unit impact parameter as
\begin{widetext}
\begin{equation}
{\cal D}_b(b;\chi_0)=
\frac{
b\displaystyle\int_0^{2\pi}
\left|\widehat I_{\Sch}(b,\alpha;\chi_0)
-\widehat I_{\rm R}(b,\alpha;\chi_0)\right|
\,\dd\alpha
}{
\displaystyle\frac12
\int_0^{2\pi}\!\int_0^{b_{\max}}
\left[\widehat I_{\Sch}(b',\alpha;\chi_0)
+\widehat I_{\rm R}(b',\alpha;\chi_0)\right]
b'\,\dd b'\,\dd\alpha
},
\qquad
D_{\rm phys}(\chi_0)=
\int_0^{b_{\max}}{\cal D}_b(b;\chi_0)\,\dd b.
\label{eq:db}
\end{equation}
\end{widetext}
Here $b_{\max}=12M$ specifies the radius of the common comparison aperture. The denominator is the same as in Eq.~\eqref{eq:dphys} and is independent of $b$ at fixed opacity. Accordingly, ${\cal D}_b$ has dimensions of inverse length and quantifies the contribution to $D_{\rm phys}$ per unit impact parameter.

Although $D_{\rm phys}$ summarizes the overall image difference, it does not indicate whether the local contrast increases or decreases in a particular region. For example, in the third column of Fig.~\ref{fig:baseline}, at $\chi_0M\simeq0.717$, the Schwarzschild and Bardeen radial profiles exhibit a larger difference in the broad-ring region $6\lesssim b/M\lesssim8$ than in the optically thin limit. Nevertheless, $D_{\rm phys}$ decreases from approximately $0.1304$ to $0.0950$. This illustrates how increased differences in one region can be outweighed by reduced contributions elsewhere. Appendix~\ref{app:inner-radius} further examines the distinction between local and integrated image differences.

To assess whether the image differences persist after accounting for the different critical-curve scales, we introduce a second measure based on the dimensionless impact parameter $\tilde b=b/b_{\rm c}$. Each image is evaluated using its own critical impact parameter, giving
\begin{widetext}
\begin{equation}
D_{\rm shape}(\chi_0)=
\frac{\displaystyle\int_{\mathcal A_{\tilde b}}
\left|
\widehat I_{\Sch}(b_{\rm c}^{\Sch}\tilde b,\alpha;\chi_0)
-\widehat I_{\rm R}(b_{\rm c}^{\rm R}\tilde b,\alpha;\chi_0)
\right|
\tilde b\,\dd\tilde b\,\dd\alpha}
{\displaystyle\frac12\int_{\mathcal A_{\tilde b}}
\left[
\widehat I_{\Sch}(b_{\rm c}^{\Sch}\tilde b,\alpha;\chi_0)
+\widehat I_{\rm R}(b_{\rm c}^{\rm R}\tilde b,\alpha;\chi_0)
\right]
\tilde b\,\dd\tilde b\,\dd\alpha}.
\label{eq:dshape}
\end{equation}
\end{widetext}
This rescaling aligns the critical curves and removes their relative radial scale difference. However, other metric-dependent effects, including those on photon propagation, redshift, and the prescribed accretion flow, remain. Thus, $D_{\rm shape}$ characterizes differences between the rescaled images rather than isolating a purely geometric or accretion-flow contribution. 

To quantify the maximum enhancement relative to the optically thin limit, we define
\begin{equation}
\Delta D=
\max_{\chi_0\in\mathcal S}
\left[D(\chi_0)-D(0)\right],
\label{eq:gain}
\end{equation}
where $D$ denotes either $D_{\rm phys}$ or $D_{\rm shape}$, and $\mathcal S$ is the set of sampled absorption coefficients, including $\chi_0=0$. By construction, $\Delta D\geq0$, with $\Delta D=0$ indicating that no enhancement is found over the sampled opacity range. When $D(0)$ is sufficiently well resolved, we also report the relative enhancement $\Delta D/D(0)$.

\subsection{Peak trajectories and the enhancement interval}

\begin{figure*}[t!]
 \centering
 \includegraphics[width=0.96\textwidth]{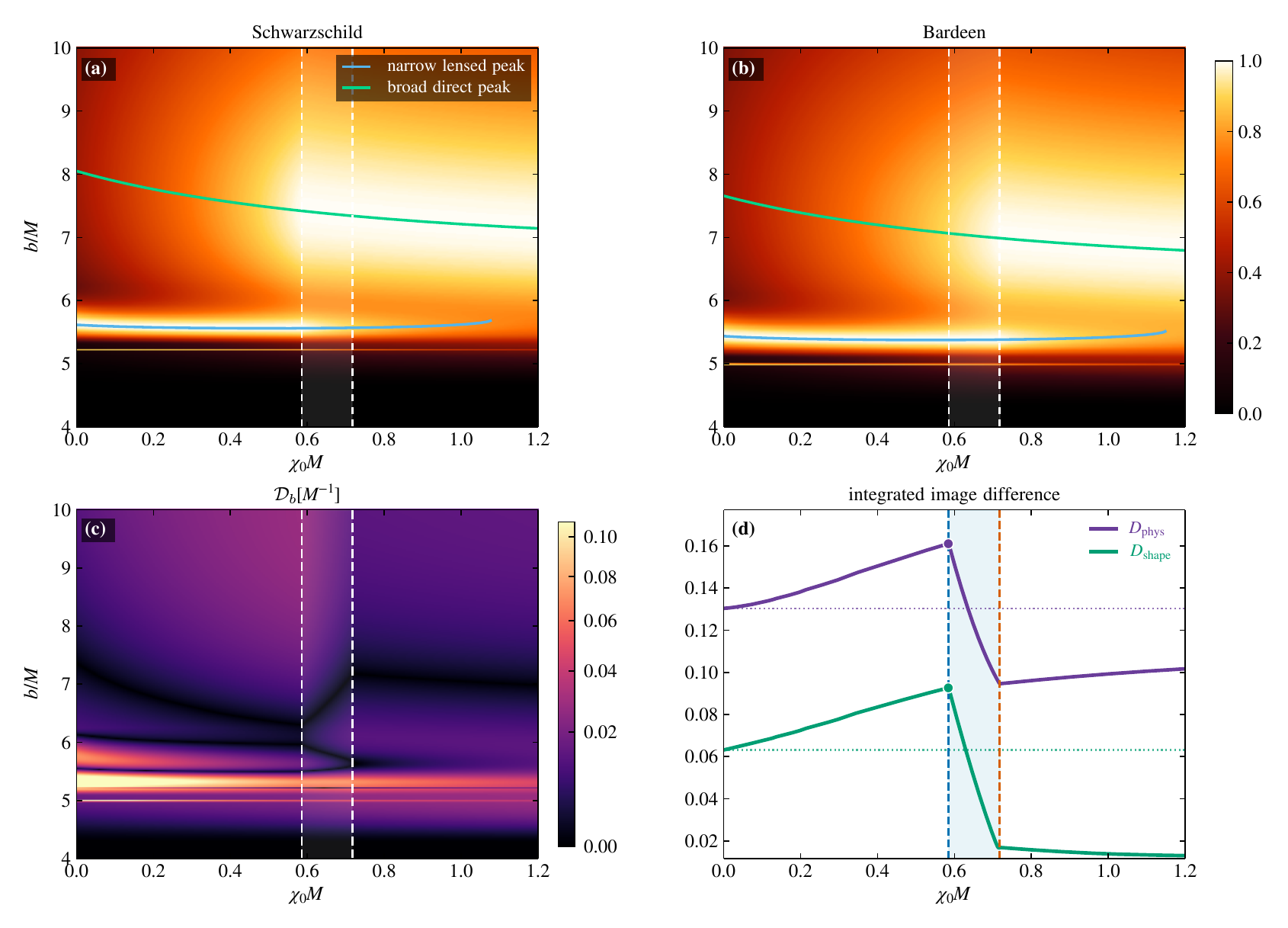}
 \caption{Evolution of the normalized intensity profiles and image differences over $0\leq\chi_0M\leq1.2$, using the same parameters as Fig.~\ref{fig:baseline}.
 Panels (a) and (b) show the radial intensity profiles $\widehat I(b;\chi_0)$ for Schwarzschild and Bardeen, respectively. The cyan and green curves trace the local maxima of the narrow lensed and broad direct-emission features. The broad peaks move inward with increasing absorption, while the narrow peaks remain identifiable beyond their respective brightness transitions.
 Panel (c) shows the radial contribution ${\cal D}_b$ to the physical-grid-integrated difference $D_{\rm phys}$.
 Panel (d) shows $D_{\rm phys}$ and the critical-curve-rescaled $D_{\rm shape}$ as functions of opacity. Their optically thin values are $0.1304$ and $0.0632$, respectively, and their maxima reach $0.1611$ and $0.0926$ near $\chi_0M\simeq0.585$, corresponding to enhancements of $23.5\%$ and $46.6\%$. The dotted horizontal lines indicate the optically thin values, while the dashed vertical lines mark the Schwarzschild and Bardeen peak transitions at $\chi_0M\simeq0.585$ and $0.717$, respectively. The shaded interval denotes the range of different dominant brightness features, within which both integrated differences decrease sharply.}
 \label{fig:trajectories}
\end{figure*}

Figure~\ref{fig:trajectories} shows the evolution of the normalized radial intensity profiles and image differences over $0\leq\chi_0M\leq1.2$. In panels (a) and (b), we track genuine local maxima within $1.02\leq b/b_{\rm c}\leq1.15$ for the narrow lensed feature and $1.28\leq b/b_{\rm c}\leq1.85$ for the broad direct-emission feature. Each trajectory is terminated when no local maximum remains within its search interval, rather than being continued along the interval boundary. As absorption increases, the broad peaks move inward, from $b\simeq8.05M$ to $7.14M$ in Schwarzschild and from $7.66M$ to $6.79M$ in Bardeen. The narrow peaks remain resolved beyond their respective brightness exchanges, with the last resolved maxima occurring at $\chi_0M\simeq1.076$ and $1.148$, respectively. Thus, the exchange of peak dominance and the disappearance of a local maximum are distinct events.

Figure~\ref{fig:trajectories}(c) shows how different radial regions contribute to $D_{\rm phys}$. The strongest localized contribution is concentrated near the narrow lensed features at $b/M\simeq5.2$--$5.5$, while weaker differences extend over a broader range of larger impact parameters. Notably, between the Schwarzschild and Bardeen peak transitions, $0.585<\chi_0M<0.717$, the radial difference distribution undergoes a pronounced rearrangement. Several low-contribution bands shift with increasing absorption, including one extending toward $b/M\simeq7$ as the Bardeen transition is approached. These dark bands correspond to radii where the two normalized intensities are nearly equal, rather than to the trajectories of the brightness maxima. The evolution therefore reflects a redistribution of the image differences across radial regions, rather than a uniform change in contrast.

The resulting behavior of the integrated measures is shown in Fig.~\ref{fig:trajectories}(d). Starting from their optically thin values, $D_{\rm phys}$ increases from $0.1304$ to a maximum of $0.1611$, while $D_{\rm shape}$ increases from $0.0632$ to $0.0926$. Both maxima occur near $\chi_0M\simeq0.585$, coinciding with the Schwarzschild brightness-peak transition. These correspond to relative enhancements of $23.5\%$ and $46.6\%$, respectively.

Interestingly, both measures subsequently decrease sharply over $0.585<\chi_0M<0.717$, despite the different dominant brightness features in the two spacetimes. By the Bardeen transition, $D_{\rm phys}$ has fallen to approximately $0.0950$, below its optically thin value, while $D_{\rm shape}$ has decreased even more substantially. This behavior illustrates that different dominant peaks do not necessarily imply a larger integrated image difference. Although the broad peak dominates Schwarzschild while the narrow peak still dominates Bardeen, the two normalized profiles can become more similar over substantial portions of the image.

\begin{figure*}[t!]
 \centering
 \includegraphics[width=0.99\textwidth]{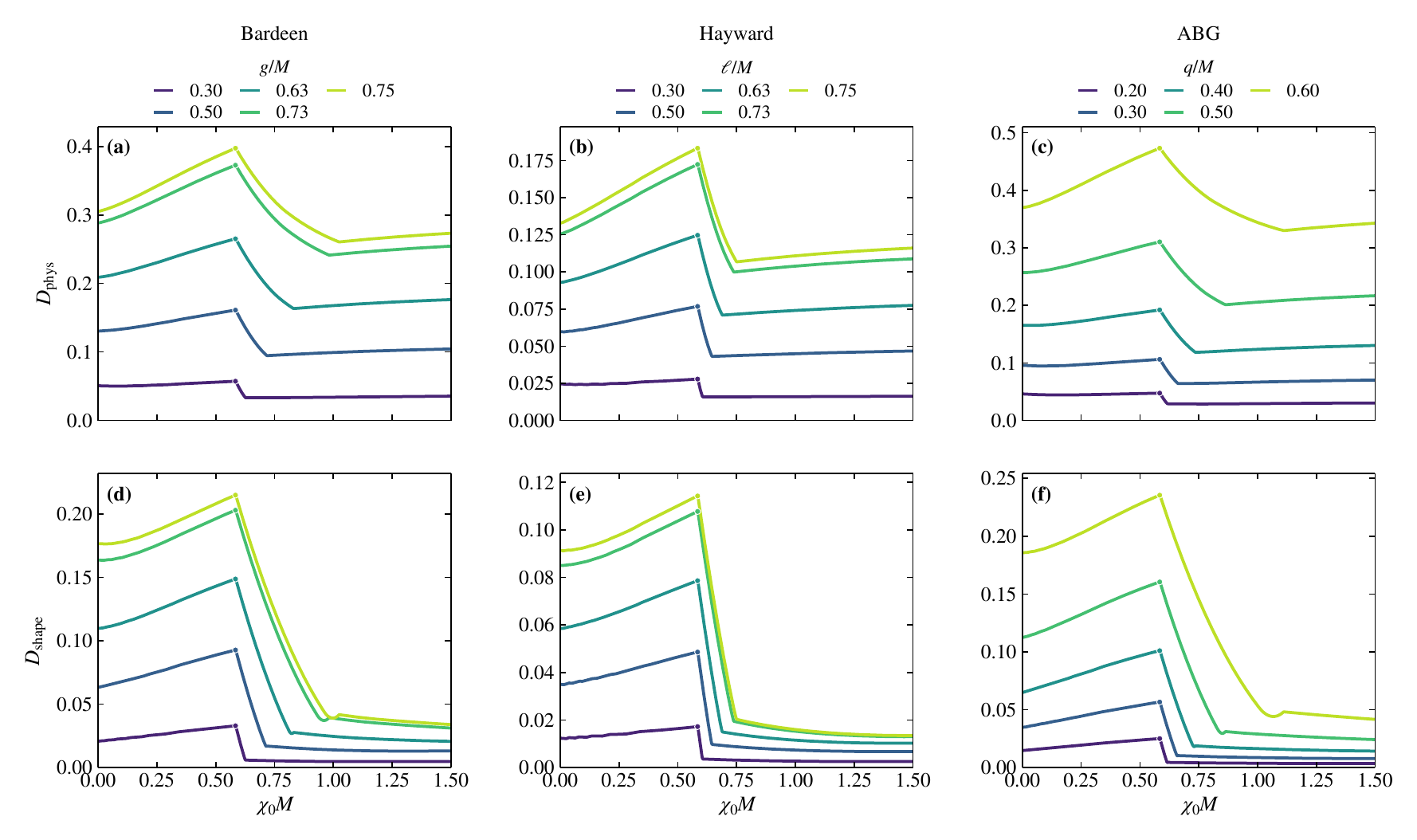}
 \caption{Dependence of the image-difference measures on effective opacity for Schwarzschild compared with Bardeen, Hayward, and ABG black holes. Panels (a--c) show the physical-grid difference $D_{\rm phys}$, while panels (d--f) show the critical-curve-rescaled difference $D_{\rm shape}$. The accretion-flow parameters are fixed at $\theta_O=0^\circ$, $\psi_0=10^\circ$, and $(\kappa_{\rm ff},\kappa_K)=(0.1,0.9)$, with the metric-dependent inner boundary $\rin=\rmbo$. Each image is independently normalized to its peak intensity before calculating either measure. Circles mark the maxima of the respective curves. For all sampled configurations, both measures attain their maxima near $\chi_0M\simeq0.585$, approximately coinciding with the Schwarzschild brightness-peak transition.  Note that the ordinate ranges differ between panels, and equal numerical values of $g/M$, $\ell/M$, and $q/M$ do not correspond to equivalent geometric deviations from Schwarzschild.}
 \label{fig:metric-families}
\end{figure*}

\begin{figure*}[t!]
 \centering
 \includegraphics[width=0.98\textwidth]{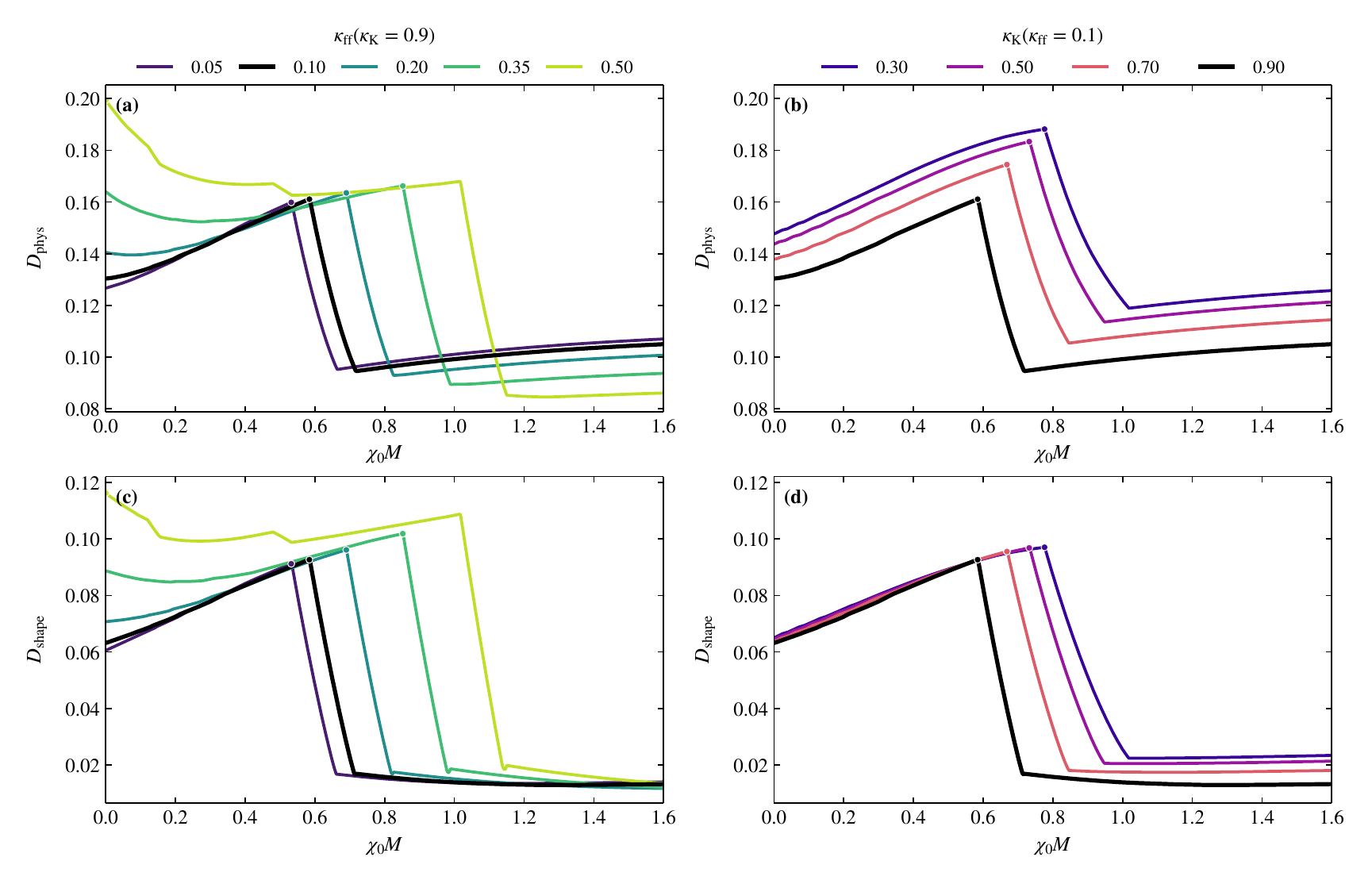}
 \caption{Dependence of the integrated image differences on the prescribed accretion-flow velocity for Schwarzschild and Bardeen ($g/M=0.5$). The other parameters are fixed at $\theta_O=0^\circ$, $\psi_0=10^\circ$, and $\rin=\rmbo$ for each spacetime. Panels (a,c) show $D_{\rm phys}$ and $D_{\rm shape}$, respectively, for varying radial-infall coefficient $\kappa_{\rm ff}$ at fixed $\kappa_K=0.9$. Panels (b,d) show the corresponding measures for varying azimuthal-velocity coefficient $\kappa_K$ at fixed $\kappa_{\rm ff}=0.1$. Colors indicate the parameter values specified in the legends, with black denoting the baseline configuration $(\kappa_{\rm ff},\kappa_K)=(0.1,0.9)$.  Increasing $\kappa_{\rm ff}$ generally suppresses the absorption-induced enhancement; at $\kappa_{\rm ff}=0.50$, neither measure exceeds its optically thin value. In contrast, positive enhancements persist throughout the displayed $\kappa_K$ scan.}
 \label{fig:velocity-robustness}
\end{figure*}

\begin{figure*}[t!]
 \centering
 \includegraphics[width=0.98\textwidth]{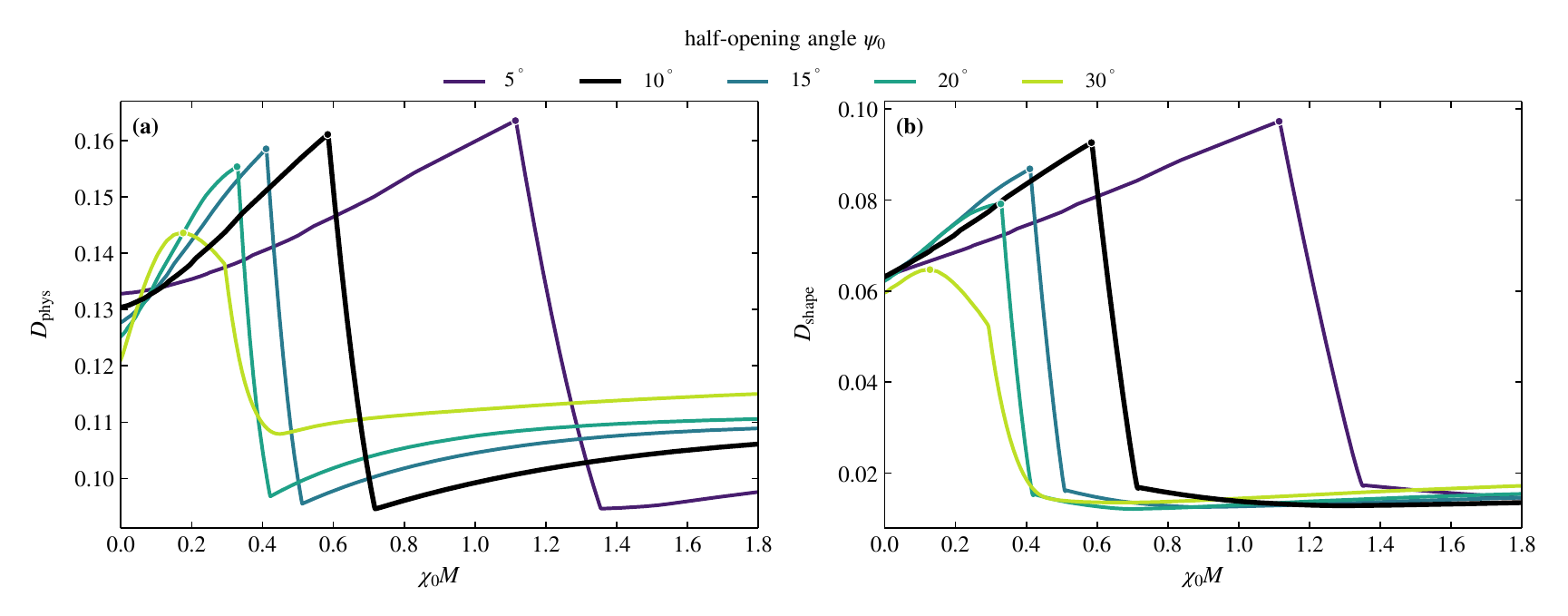}
 \caption{Dependence of the integrated image differences on the accretion-flow half-opening angle $\psi_0$ for the face-on Schwarzschild--Bardeen comparison with $g/M=0.5$. Panels (a) and (b) show $D_{\rm phys}$ and $D_{\rm shape}$, respectively, for $\psi_0=5^\circ$, $10^\circ$, $15^\circ$, $20^\circ$, and $30^\circ$, with the remaining parameters fixed at their baseline values.  Circles mark the maxima over the sampled opacity range. Positive enhancements are found at all displayed opening angles. Increasing the flow thickness generally shifts the maxima toward lower opacities and suppresses the shape enhancement more strongly than the physical-grid enhancement.}
 \label{fig:opening-angle}
\end{figure*}

\begin{figure*}[t!]
 \centering
 \includegraphics[width=0.98\textwidth]{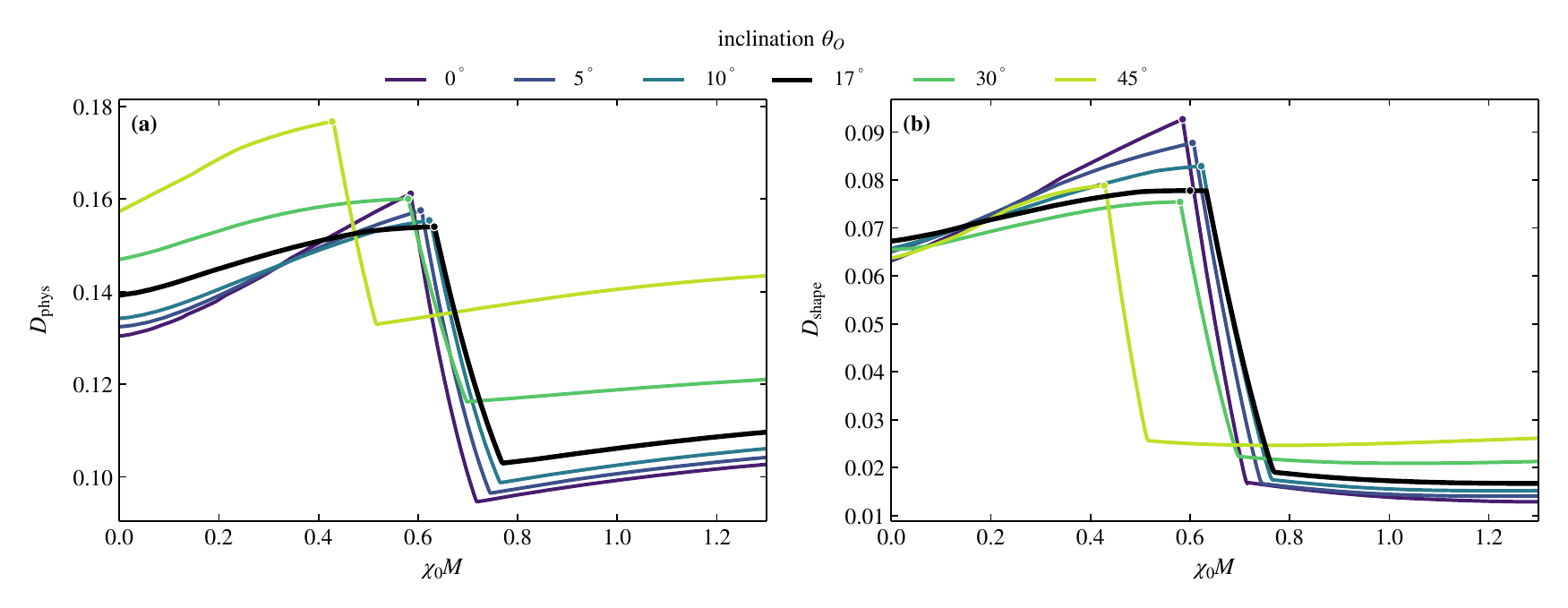}
 \caption{Dependence of the integrated image differences on observer inclination $\theta_O$ for the Schwarzschild--Bardeen comparison with $g/M=0.5$, keeping the remaining parameters fixed at their baseline values. Panels (a) and (b) show $D_{\rm phys}$ and $D_{\rm shape}$, respectively, evaluated from the full two-dimensional images. The displayed inclinations are $\theta_O=0^\circ$, $5^\circ$, $10^\circ$, $17^\circ$, $30^\circ$, and $45^\circ$. Colors indicate the inclinations specified in the legend, with black denoting the illustrative M87*-motivated angle $\theta_O=17^\circ$. Circles mark the maxima over the sampled opacity range.}
 \label{fig:inclination}
\end{figure*}

The enhancement of $D_{\rm shape}$ shows that the effect is not solely attributable to the difference in critical-curve scales, although the rescaling does not eliminate other metric-dependent effects on the images. As illustrated by Fig.~\ref{fig:trajectories}(c), the integrated measures do not capture every local change in image contrast. Nevertheless, $D_{\rm phys}$ and $D_{\rm shape}$ provide concise global diagnostics for systematic comparisons. We therefore use them in the following section to examine the dependence of the absorption-induced enhancement on the black-hole geometry and accretion-flow parameters.

\section{Dependence on geometry and flow parameters}
\label{sec:conditions}

Having established the absorption-induced enhancement in the baseline Schwarzschild--Bardeen comparison, we now examine its dependence on black-hole geometry and accretion-flow parameters. We retain the emissivity prescription in Eq.~\eqref{eq:emissivity} and vary one model parameter at a time, keeping the remaining settings fixed at their baseline values. Using $D_{\rm phys}$ and $D_{\rm shape}$ as our primary diagnostics, we investigate the magnitude and parameter dependence of the enhancement, without assuming that the transition opacity or enhancement is universal.

\subsection{Regular-black-hole families}
\label{sec:families}

We first examine the dependence of the absorption-induced enhancement on the black-hole geometry by comparing Schwarzschild with the Bardeen, Hayward, and ABG spacetimes. For Bardeen and Hayward, we sample $g/M$ and $\ell/M$ at $(0.30,0.50,0.63,0.73,0.75)$, respectively, while for ABG we use $q/M=(0.20,0.30,0.40,0.50,0.60)$. All selected configurations possess an outer horizon, with the ABG charges remaining below the critical value $|q|/M\simeq0.63418$. The accretion-flow parameters are fixed at their baseline values, with $\rin=\rmbo$ evaluated separately for each spacetime.

Figure~\ref{fig:metric-families} shows that all three families exhibit positive enhancements in both $D_{\rm phys}$ and $D_{\rm shape}$ over the sampled parameter ranges. The maxima consistently occur near $\chi_0M\simeq0.585$, approximately coinciding with the brightness-peak transition of the common Schwarzschild reference. This suggests that the evolution of the reference image plays an important role in determining the opacity at which the largest differences arise. However, the enhancement amplitudes depend strongly on the regular-black-hole geometry.

For Bardeen, the maximum $D_{\rm phys}$ increases from $0.0573$ at $g/M=0.30$ to $0.3974$ at $g/M=0.75$. The corresponding values for Hayward increase from $0.0279$ to $0.1833$ as $\ell/M$ varies over the same range. At the largest sampled parameters, the relative physical-grid enhancements reach $30.3\%$ and $38.1\%$, respectively. Positive enhancements in $D_{\rm shape}$ are also found throughout these scans, demonstrating that the effect persists after matching the critical-curve scales.

The ABG family exhibits a similar enhancement, with its brightness-peak transition occurring after that of Schwarzschild. The ABG transition opacity increases from $\chi_{\rm tr}M\simeq0.617$ at $q/M=0.20$ to $1.111$ at $q/M=0.60$. Over the same charge range, the maximum $D_{\rm phys}$ increases from $0.0475$ to $0.4731$. At $q/M=0.60$, $D_{\rm phys}$ rises from its optically thin value of $0.3697$ to $0.4731$, while $D_{\rm shape}$ increases from $0.1857$ to $0.2355$, corresponding to relative enhancements of approximately $28.0\%$ and $26.8\%$, respectively. In contrast, at $q/M=0.20$, the physical-grid enhancement is only about $2.9\%$, whereas the shape enhancement reaches approximately $70.9\%$ relative to its smaller optically thin baseline. These results illustrate that the relative enhancements of the two measures need not follow the same trend.


\subsection{Flow velocity}
\label{sec:velocity}

We next examine the dependence of the absorption-induced enhancement on the prescribed flow velocity. We vary the radial-infall coefficient $\kappa_{\rm ff}=(0.05,0.10,0.20,0.35,0.50)$ at fixed $\kappa_K=0.9$, and the azimuthal-velocity coefficient $\kappa_K=(0.30,0.50,0.70,0.90)$ at fixed $\kappa_{\rm ff}=0.1$. Both parameters affect the velocity-dependent redshift factor. Although the emissivity scales as $j(r)\propto\kappa_K^2$ when $\kappa_K$ is varied, this overall factor cancels under peak normalization.

Figures~\ref{fig:velocity-robustness}(a) and (c) show that increasing $\kappa_{\rm ff}$ progressively weakens the absorption-induced enhancement, even though the optically thin image differences generally increase. At $\kappa_{\rm ff}=0.05$, the maximum increase in $D_{\rm phys}$ relative to its thin-state value is $0.0332$. In contrast, at $\kappa_{\rm ff}=0.50$, neither $D_{\rm phys}$ nor $D_{\rm shape}$ exceeds its optically thin value over the sampled opacity range. Nevertheless, the brightness-peak transitions in Schwarzschild and Bardeen remain separated by approximately $0.13$ in $\chi_0M$ throughout the scan. This demonstrates that asynchronous peak exchanges alone do not guarantee an enhancement of the integrated image difference.

In contrast, positive enhancements persist throughout the azimuthal-velocity scan, as shown in Figs.~\ref{fig:velocity-robustness}(b) and (d). Decreasing $\kappa_K$ from $0.9$ to $0.3$ shifts the maxima toward larger absorption coefficients and increases the physical-grid enhancement from $0.0307$ to $0.0406$. The corresponding shape enhancements vary more modestly, between $0.0295$ and $0.0324$. Thus, within the parameter ranges considered, the absorption-induced enhancement is more sensitive to changes in radial infall, while remaining comparatively robust under variations in azimuthal motion.

\subsection{Thickness and viewing angle}
\label{sec:inclination}

We next examine the effects of accretion-flow thickness and observer inclination. Figure~\ref{fig:opening-angle} shows the dependence on the flow half-opening angle $\psi_0=(5^\circ,10^\circ,15^\circ,20^\circ,30^\circ)$ for a face-on observer, with the remaining parameters fixed at their baseline values. The overall $1/\psi_0$ factor in the emissivity cancels under peak normalization. The dependence on thickness therefore arises from changes in the emitting and absorbing geometry, including the photon path lengths within the flow, the intersected emission regions, and the latitude dependence of the redshift.

The absolute physical-grid enhancement $\Delta D_{\rm phys}$ remains close to $0.03$ for the sampled opening angles $5^\circ\leq\psi_0\leq20^\circ$ and is reduced to $0.0224$ at $\psi_0=30^\circ$. The shape enhancement is more sensitive to thickness, with $\Delta D_{\rm shape}$ decreasing from $0.0341$ at $5^\circ$ to $0.0050$ at $30^\circ$. For $\psi_0=30^\circ$, the maxima of $D_{\rm phys}$ and $D_{\rm shape}$ occur at $\chi_0M\simeq0.176$ and $0.128$, respectively, preceding the Schwarzschild and Bardeen brightness-peak transitions at $\chi_0M\simeq0.293$ and $0.454$. Thus, the largest integrated image differences need not coincide with the brightness-peak exchanges, particularly for thicker flows.

Figure~\ref{fig:inclination} examines the dependence on observer inclination, with $\theta_O=(0^\circ,5^\circ,10^\circ,17^\circ,30^\circ,45^\circ)$. For inclined observations, the image brightness becomes azimuthally asymmetric, and both $D_{\rm phys}$ and $D_{\rm shape}$ are evaluated using the full two-dimensional intensity distributions $\widehat I(b,\alpha;\chi_0)$. This retains the angular information, including Doppler-induced brightness asymmetry, that would be lost in azimuthally averaged radial profiles.

At $\theta_O=17^\circ$, an illustrative M87*-motivated inclination, $D_{\rm phys}$ increases from approximately $0.1392$ in the optically thin limit to $0.1540$, while $D_{\rm shape}$ increases from $0.0673$ to $0.0778$. These results indicate that absorption-induced enhancement can persist in asymmetric images, even though its magnitude and the opacity at which it is largest depend on the viewing geometry.

\section{Discussion and conclusions}
\label{sec:discussion}

We have shown that absorption can enhance morphological differences between black-hole images with similar critical impact parameters within a phenomenological accretion-flow model. Using the same modeling prescription, we compared Schwarzschild with Bardeen, Hayward, and Ay\'on-Beato--Garc\'ia spacetimes. In the face-on Schwarzschild--Bardeen example with $g/M=0.5$, the dominant brightness feature changes from a narrow lensed peak to a broader outer peak at $\chi_0M\simeq0.585$ and $0.717$, respectively. Near the Schwarzschild transition, the integrated difference between independently normalized images increases from $0.1304$ to $0.1611$ ($23.5\%$) on the common physical grid, and from $0.0632$ to $0.0926$ ($46.6\%$) after rescaling by the respective critical impact parameters. The persistence of this enhancement after rescaling indicates that it cannot be attributed solely to differences in the critical-curve scale.

The parameter scans show that the enhancement occurs across all three regular-black-hole families and over a broad range of the sampled flow parameters and viewing inclinations. Its magnitude and occurrence nevertheless depend on the prescribed flow, and exceptions arise for some velocity and inner-boundary configurations. In particular, for $\kappa_{\rm ff}=0.5$, neither integrated measure exceeds its optically thin value, despite the separated peak exchanges. Asynchronous peak exchanges are therefore insufficient to produce an enhancement. Moreover, a decrease in the integrated difference does not imply that all image regions become more similar: local contrasts can increase while being outweighed by reduced differences elsewhere, as demonstrated in Appendix~\ref{app:inner-radius}. These results distinguish the evolution of individual brightness features from that of the global image-difference measures.

The broader implication is that the opacity dependence of image morphology may provide information complementary to the critical-curve scale. Images of different spacetimes with similar critical impact parameters can respond differently to absorption, changing the relative prominence of their brightness features. Comparing image morphology across different opacity regimes may therefore help distinguish spacetime models whose characteristic shadow scales are similar. The present results motivate examining this possibility through the evolution of the brightness distribution, rather than relying on the apparent ring size or a single integrated difference alone.

Our treatment remains phenomenological: the flow velocity and emissivity are prescribed, and absorption is represented by a frequency-independent effective coefficient. Although the different ring sizes reported for M87* at 1.3 and 3.5~mm motivate this study~\cite{EHT2019I,Lu2023}, changing the observing frequency generally modifies both emission and absorption. Varying $\chi_0$ at fixed emissivity therefore cannot be directly identified with a change in observing frequency. Establishing whether the morphological responses found here improve observational discrimination will require physically consistent, frequency-dependent radiative transfer, potentially informed by GRMHD simulations, together with an assessment of accretion-flow uncertainties and instrumental effects. The present study provides a controlled demonstration that absorption can amplify morphological differences beyond those characterized by the critical impact parameter, and a motivation for investigating whether such differences survive in realistic multifrequency comparisons.

\section*{Acknowledgments}

This work  is supported by the National Natural Science Foundation of China  (12405063).

\appendix

\section{Optical depth and the peak exchange}
\label{app:optical-depth}

\begin{table*}[t!]
 \centering
 \caption{Total optical depths and contributions to the observed intensity at the Schwarzschild and Bardeen brightness-peak transitions. The last column gives the fraction of the observed intensity originating from locations with foreground optical depth $\tau_{\rm fg}>1$. Contributions from cells crossing $\tau_{\rm fg}=1$ are split at the threshold. The parameters correspond to the face-on baseline with Bardeen $g/M=0.5$.}
 \label{tab:optical-depth}
 \begin{tabular}{lccccc}
  \toprule
  Spacetime & $\chi_0M$ & Peak & $b/M$ & $\tau_{\rm tot}$ &
  \shortstack{Observed contribution from\\$\tau_{\rm fg}>1$ (\%)} \\
  \midrule
  Schwarzschild & $0.585$ & Narrow & $5.562$ & $3.531$ & $50.3$ \\
                &         & Broad  & $7.420$ & $1.401$ & $16.8$ \\
  Bardeen       & $0.717$ & Narrow & $5.381$ & $4.129$ & $44.8$ \\
                &         & Broad  & $6.988$ & $1.615$ & $21.8$ \\
  \bottomrule
 \end{tabular}
\end{table*}

The brightness-peak transition discussed in Sec.~\ref{sec:phenomenon} should be distinguished from the condition of unit optical depth. Along a given ray, the total optical depth accumulated within the absorbing region is
\begin{equation}
    \tau_{\rm tot}
    =\chi_0\int_{\rm ray}\dd\ell_{\rm stat}.
    \label{eq:total-ray-depth}
\end{equation}
However, the observed intensity depends on the foreground optical depth $\tau_{\rm fg}$ between each emission point and the observer, rather than on $\tau_{\rm tot}$ alone. As described by the transfer prescription in Eq.~\eqref{eq:cell}, radiation emitted at different locations along the same ray experiences different attenuation.

Table~\ref{tab:optical-depth} compares the narrow- and broad-peak rays at their respective brightness transitions. Although all four rays have $\tau_{\rm tot}>1$, their observed intensities receive markedly different contributions from regions with large foreground optical depths. Approximately $45$--$50\%$ of the narrow-peak intensity originates from locations with $\tau_{\rm fg}>1$, compared with only $17$--$22\%$ for the broad peaks. The broad features thus receive a substantially larger fraction of their observed intensity from regions with smaller foreground optical depths.

These results support the interpretation that the narrow and broad brightness features respond differently to absorption because of their distinct emission and foreground-depth distributions. The brightness exchange depends on the combined effects of emissivity, gravitational redshift, and attenuation, rather than on a universal threshold $\tau_{\rm tot}=1$.

\section{Dependence on the inner boundary}
\label{app:inner-radius}

\begin{figure*}[t!]
 \centering
 \includegraphics[width=0.96\textwidth]{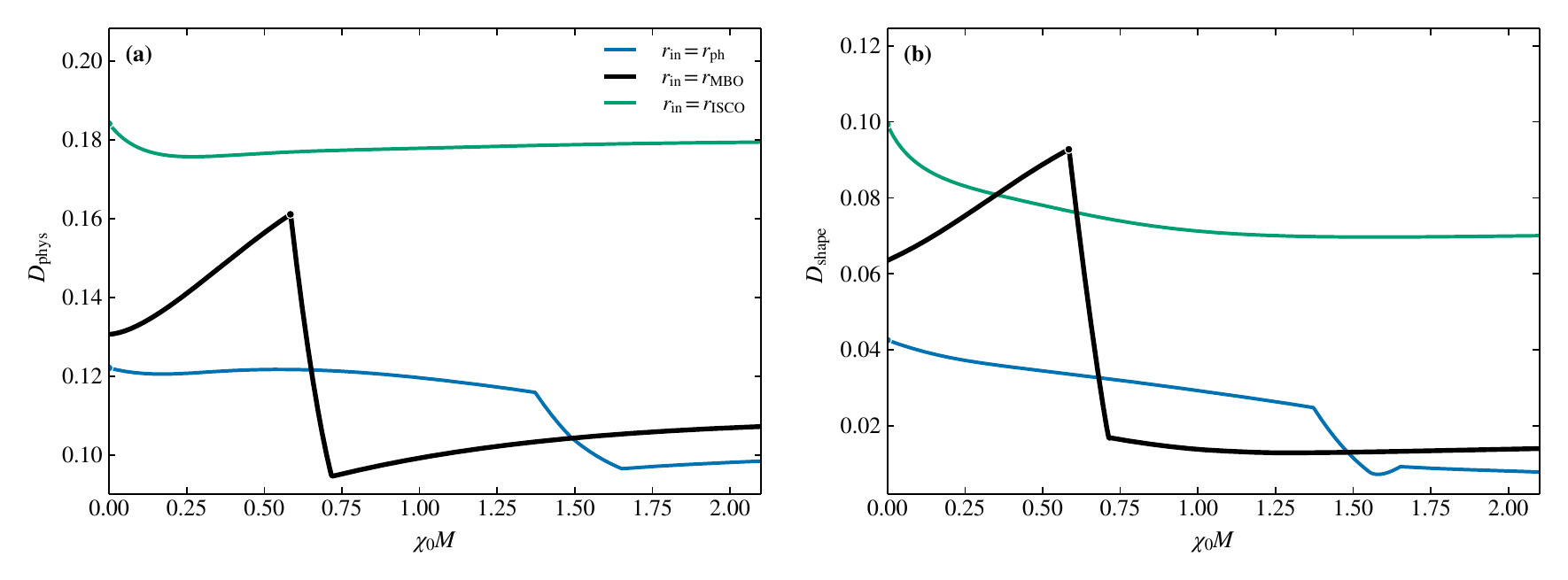}
 \caption{Dependence of the integrated image differences on the inner-boundary prescription for Schwarzschild and Bardeen with $g/M=0.5$. Panels (a) and (b) show $D_{\rm phys}$ and $D_{\rm shape}$, respectively, for the metric-dependent photon-sphere, MBO, and ISCO radii. All other parameters retain their baseline values. The black curves denote the MBO prescription, and the circles mark the sampled maxima. The different responses demonstrate the sensitivity of the absorption-induced image differences to the adopted inner boundary.}
 \label{fig:inner-edge}
\end{figure*}

\begin{figure*}[t!]
 \centering
 \includegraphics[width=0.98\textwidth]{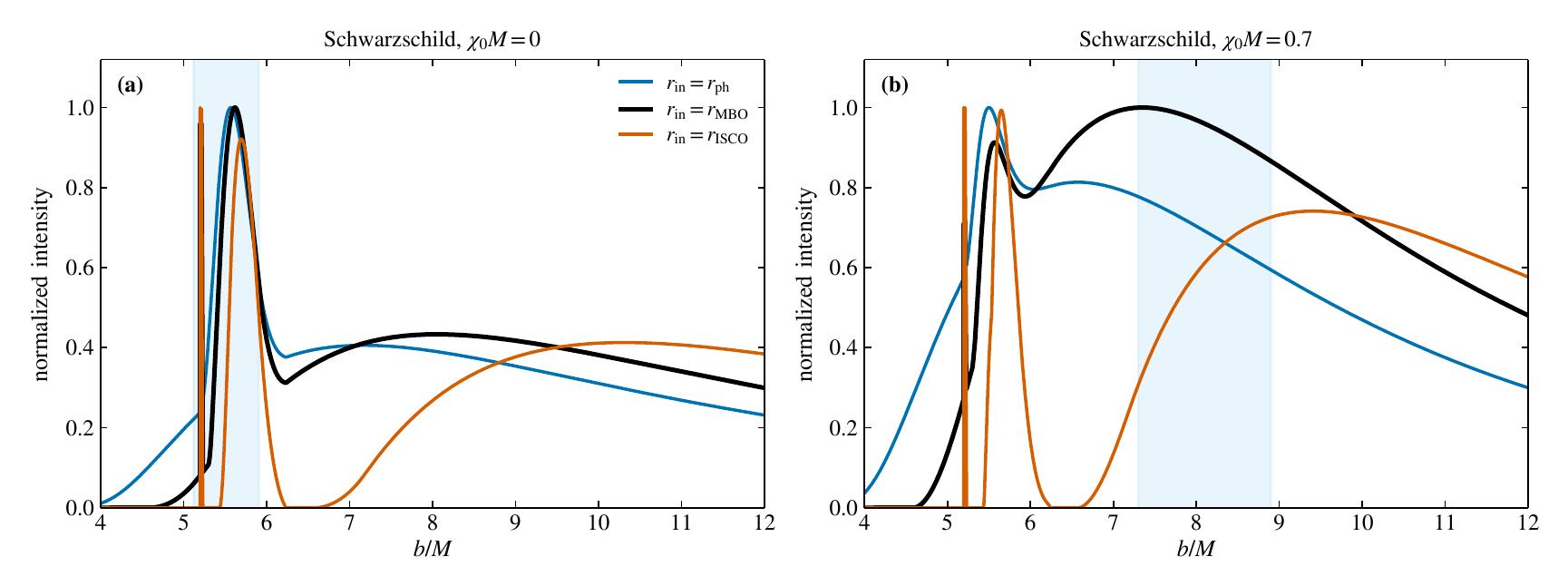}
\caption{Normalized Schwarzschild radial intensity profiles for three inner-boundary prescriptions. Panels (a) and (b) correspond to $\chi_0M=0$ and $0.7$, respectively, with the remaining model parameters fixed at their baseline values. Blue, black, and orange denote $\rin=r_{\rm ph}=3M$, $\rin=\rmbo=4M$, and $\rin=r_{\rm ISCO}=6M$. The shaded region in panel (a) indicates the illustrative radius range $5.12\leq b/M\leq5.91$ inferred from the reported $1.3\,{\rm mm}$ ring diameter of $42\pm3\,\mu\mathrm{as}$ for M87*~\cite{EHT2019I}. The shaded region in panel (b) indicates the illustrative radius range $7.3\leq b/M\leq8.9$ inferred from the reported $3.5\,{\rm mm}$ ring diameter of M87*~\cite{Lu2023}. At $\chi_0M=0.7$, the MBO profile has a dominant broad peak at $b\simeq7.35M$, within the shaded range, whereas the photon-sphere and ISCO prescriptions retain their brightest features closer to the critical curve.}
 \label{fig:inner-morphology}
\end{figure*}

\begin{table*}[t!]
 \centering
 \caption{Location $b/M$ of the brightest peak in Schwarzschild images for three inner-boundary prescriptions at selected effective opacities $\chi_0M$.}
 \label{tab:inner-morphology}
 \begin{tabular}{lccccc}
  \toprule
  Inner radius & $0$ & $0.6$ & $1$ & $4$ & $30$ \\
  \midrule
  $r_{\rm ph}=3M$   & $5.57$ & $5.50$ & $5.51$ & $6.10$ & $6.01$ \\
  $r_{\rm MBO}=4M$  & $5.62$ & $7.41$ & $7.21$ & $6.88$ & $6.80$ \\
  $r_{\rm ISCO}=6M$ & $5.20$ & $5.20$ & $5.20$ & $5.20$ & $5.20$ \\
  \bottomrule
 \end{tabular}
\end{table*}

\begin{figure*}[t!]
 \centering
 \includegraphics[width=\textwidth]{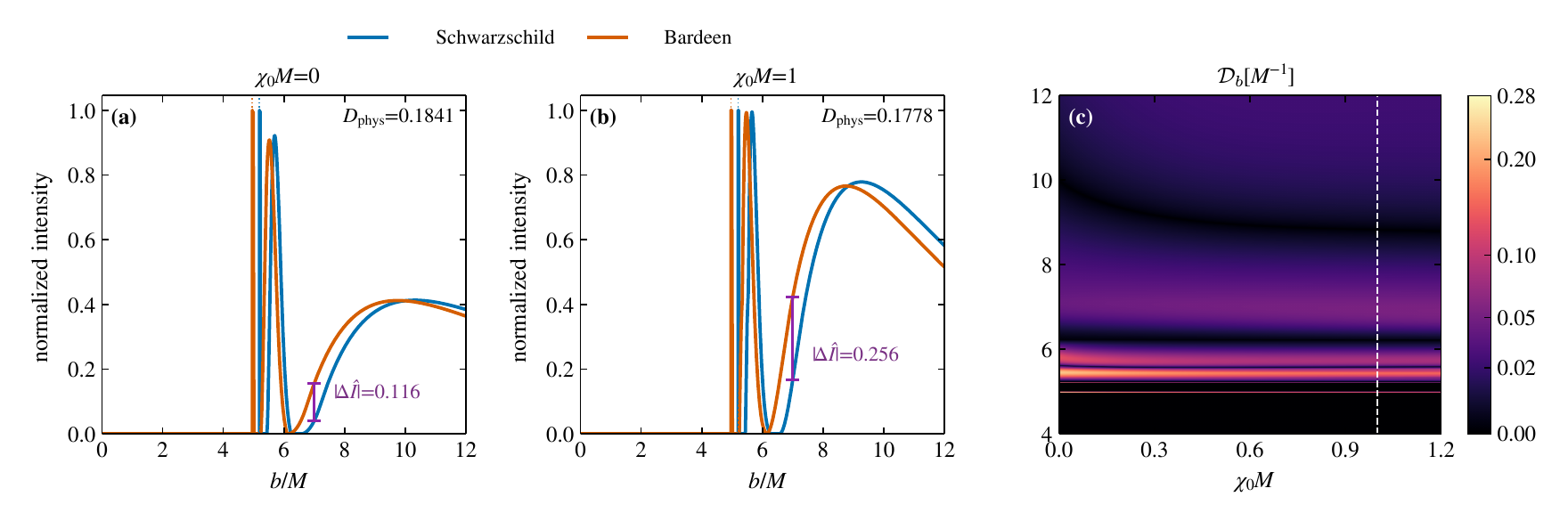}
 \caption{Local and integrated image differences for the ISCO inner-boundary prescription, with $g/M=0.5$ and the other parameters fixed at their baseline values. Each spacetime uses its own ISCO, $\rin/M=6$ for Schwarzschild and $5.5722$ for Bardeen. Panels (a) and (b) compare their normalized radial profiles at $\chi_0M=0$ and $1$, respectively; the purple segments indicate the absolute local intensity difference at $b=7M$. The corresponding integrated differences $D_{\rm phys}$ are displayed in the panels. Panel (c) shows the radial contribution ${\cal D}_b(b;\chi_0)$ to $D_{\rm phys}$ as a function of impact parameter and opacity, with the dashed vertical line marking $\chi_0M=1$. The integrated difference decreases even though the local difference at $b=7M$ increases.}
 \label{fig:inner-local}
\end{figure*}

The inner radius $\rin$ enters the phenomenological flow model in two ways: it sets the inner boundary of the emitting and absorbing region and also appears in the emissivity prescription in Eq.~\eqref{eq:emissivity}. Its choice can therefore affect both the image morphology and the absorption-induced differences between spacetimes. Here we compare three prescriptions, $\rin=r_{\rm ph}$, $\rin=\rmbo$, and $\rin=r_{\rm ISCO}$, with each radius evaluated in its respective metric. We use the Schwarzschild--Bardeen pair with $g/M=0.5$ and keep $\theta_O=0^\circ$, $\psi_0=10^\circ$, and $(\kappa_{\rm ff},\kappa_K)=(0.1,0.9)$ fixed. The comparison changes the full inner-boundary and emissivity prescription, rather than the location of a geometric cutoff alone.

Figure~\ref{fig:inner-edge} shows that the integrated image differences depend sensitively on the inner-boundary prescription. The MBO choice produces a clear intermediate-opacity enhancement in both $D_{\rm phys}$ and $D_{\rm shape}$, whereas the photon-sphere and ISCO choices do not exhibit a comparable enhancement in both measures. Notably, although the ISCO prescription yields larger image differences in the optically thin limit, neither measure exceeds its corresponding thin-state value as absorption increases. These results highlight the dependence of the enhancement on the adopted accretion-flow prescription, rather than on the background geometry alone.

The MBO prescription is adopted not to maximize the image-difference enhancement, but on morphological grounds. Within the same phenomenological flow model, it allows a narrow bright feature near the critical curve in the optically thin limit and a broader, larger-radius feature at finite absorption. This behavior is qualitatively motivated by the different ring sizes observed in M87* at $1.3$ and $3.5\,{\rm mm}$~\cite{Lu2023}. As discussed in Sec.~\ref{sec:model}, the $3.5\,{\rm mm}$ observation corresponds to an apparent ring-radius range of $7.3M$--$8.9M$, with the larger ring interpreted in accretion-flow models as arising from optically thick emission due to absorption.

To examine this morphological motivation, we first consider Schwarzschild, for which $r_{\rm ph}=3M$, $\rmbo=4M$, and $r_{\rm ISCO}=6M$. Figure~\ref{fig:inner-morphology} compares the normalized radial intensity profiles at $\chi_0M=0$ and $0.7$, while Table~\ref{tab:inner-morphology} tracks the location of the brightest feature at selected opacities. The global maximum is identified over $0\leq b/M\leq20$, without restricting the search to predefined narrow- or broad-peak intervals. In the optically thin limit, all three prescriptions produce a dominant narrow feature close to $b_{\rm c}$, so the thin-state morphology alone does not distinguish the MBO choice.

Their behavior at finite absorption, however, differs substantially. For $\rin=\rmbo$, the dominant broad peak lies at $b\simeq7.35M$ for $\chi_0M=0.7$, within the illustrative observational range $7.3M$--$8.9M$. The total optical depth along this peak ray is $\tau_{\rm tot}\simeq1.66$, indicating appreciable absorption. More generally, the dominant MBO peak lies between $7M$ and $8M$ over the sampled interval $0.59\lesssim\chi_0M\lesssim1.91$, overlapping the observational reference range for approximately $0.59\lesssim\chi_0M\lesssim0.79$. In contrast, neither the photon-sphere nor the ISCO prescription produces a dominant peak within this range over $0\leq\chi_0M\leq4$. The photon-sphere prescription retains its brightest feature at smaller impact parameters, whereas the ISCO prescription remains dominated by narrow features close to $b_{\rm c}$.

Finally, the ISCO prescription provides a useful example of how integrated image differences can evolve differently from local contrast. As shown in Figs.~\ref{fig:inner-local}(a) and (b), increasing the opacity from $\chi_0M=0$ to $1$ reduces $D_{\rm phys}$ from $0.184$ to $0.178$, while the local absolute intensity difference $|\widehat I_{\Sch}-\widehat I_{\B}|$ at $b=7M$ increases from $0.116$ to $0.256$. The radial contribution ${\cal D}_b$ in Fig.~\ref{fig:inner-local}(c) further illustrates the nonuniform evolution of the image differences. Using the area weighting and normalization in Eq.~\eqref{eq:db}, the contribution from $b/M\geq6.2$ increases from $0.088$ to $0.119$, whereas that from $b/M<6.2$ decreases from $0.096$ to $0.059$.  Thus, increased differences in one radial region can be outweighed by reduced contributions elsewhere, leading to a smaller integrated difference despite stronger local contrast.

\bibliographystyle{jhep}
\bibliography{references_v21}
\end{document}